\documentclass{JHEP}

\usepackage{graphicx}
\usepackage{psfrag}
\usepackage{url}

\providecommand{\GeV}{{\rm GeV}}
\providecommand{\MSbar }{\ensuremath{ \overline{\rm MS} }}

\makeatletter 
\preprint{hep-ph/0204127\\
\@date
}
\makeatother

\title{Parton Distribution Functions suitable for Monte-Carlo
        event generators}

\author{John C. Collins, Xiaomin Zu\\
        Physics Department,
        Penn State University,\\
        104 Davey Laboratory,
        University Park PA 16802
        U.S.A.
}

\date{
      13 June 2002}

\abstract{In the usual factorization theorems, which give predictions
  only for inclusive cross sections, there is considerable freedom in
  the choice of the scheme to define the parton distribution functions.  These
  theorems do not directly apply to Monte-Carlo event generators, and
  more general factorization theorems which give predictions for fully
  exclusive cross sections are needed.  It has been shown that
  appropriate parton distribution functions are uniquely defined by the 
  showering
  algorithm. In this paper, we present results of calculations of the
  Monte-Carlo parton distribution functions in terms of the commonly used 
  \MSbar{}  parton distribution functions.  At small $x$ the differences are 
  large, which  demonstrates the importance of using 
  the correct parton distribution functions in
  an event generator rather than \MSbar{} parton distribution functions.  
  We present some
  simple approximations that enable an understanding of the sizes of
  the results to be obtained.
}


\begin{document}
\section{Introduction}
\label{sec:intro}
The standard factorization theorems \cite{handbook} of QCD only give
predictions for inclusive cross sections. In Monte-Carlo (MC) event
generators we generate complete events and implement QCD predictions
for the detailed structure of the final state, therefore it is
important to get both the inclusive cross section and the exclusive
cross sections right.

For inclusive cross sections, there is freedom in choosing  the
factorization scheme that defines the parton distribution functions
(pdfs).  But one of us has shown \cite{JCC} that this is not the case in MC
event generators; the specific showering algorithm used in a
particular event generator entails a particular definition of the pdfs
that should be used in the event generator.\footnote{
   See the Note Added at the end of the paper for earlier work on the
   same idea.
}
In this paper we first
expand these arguments, showing how they are related to the
requirement of obtaining correct exclusive cross sections.  

We then present and analyze the results of calculations for pdfs
 that are appropriate to the algorithm of Bengtsson,
Sj{\"o}strand and van Zijl \cite{BS}, as is used, for example, in the
event generators PYTHIA and RAPGAP.  In order to reach the
next-to-leading order (NLO) accuracy in both the inclusive and the
exclusive cross sections, it is important to use the correct pdfs and
the correspondingly determined NLO hard scattering coefficients.  As
an example, we show that in the DIS $F_2$ calculation, using \MSbar{} pdfs
instead of the correct ones for the specific event generator
introduces an error of roughly $40\%$ at small $x$.  This can, of
course, substantially affect the phenomenology.

\section{Factorization Schemes and Parton Distribution Functions in 
Monte-Carlo event generators}
\label{sec:majour}

\subsection{Factorization theorem and Monte-Carlo event generators}
\label{sec:scheme}

The factorization theorem states that appropriate inclusive cross
sections with a large transverse momentum $Q$ are given
\cite{handbook} (to the leading power in $Q$) by a hard scattering
coefficient convoluted with pdfs.  Each hard
scattering coefficient is infrared safe, calculable in perturbation
theory and independent of the external hadron.  The pdfs contain all
the infrared sensitivity of the original cross section, they are
external-hadron specific and are independent of the particular hard
scattering process.  

For inclusive cross sections, it is well known that there is some freedom
in choosing the prescription by which the pdfs are defined.  A set of
rules that makes the choices is called a `factorization scheme'.  Such
a scheme both defines the pdfs and implies a rule for unambiguously
calculating the hard scattering coefficients.

It might be concluded that this also applies to the hard scattering
coefficients in an MC event generator.  As shown by one of us
\cite{JCC}, this is not in fact the case, and we will now review the
reasoning.\footnote{This implies that we disagree with the reasoning
  about NLO corrections to event generators that assumes an opposite
  conclusion, as in Refs.\ \cite{Dobbs,Poetter}.
}

An MC event generator calculates the exclusive components of the cross
section, by using a combination of perturbatively calculated
quantities for the larger scales and suitable modeling for the
nonperturbative physics.  The perturbative part consists of
hard-scattering coefficients and evolution kernels.  For the idea of
computing NLO (and even higher order) corrections to make practical
sense, the perturbative expansions of the hard scattering coefficients
and the evolution kernels must be free of logarithms of large ratios
of kinematic variables.

\FIGURE{
   \centering
   \includegraphics[scale=0.5]{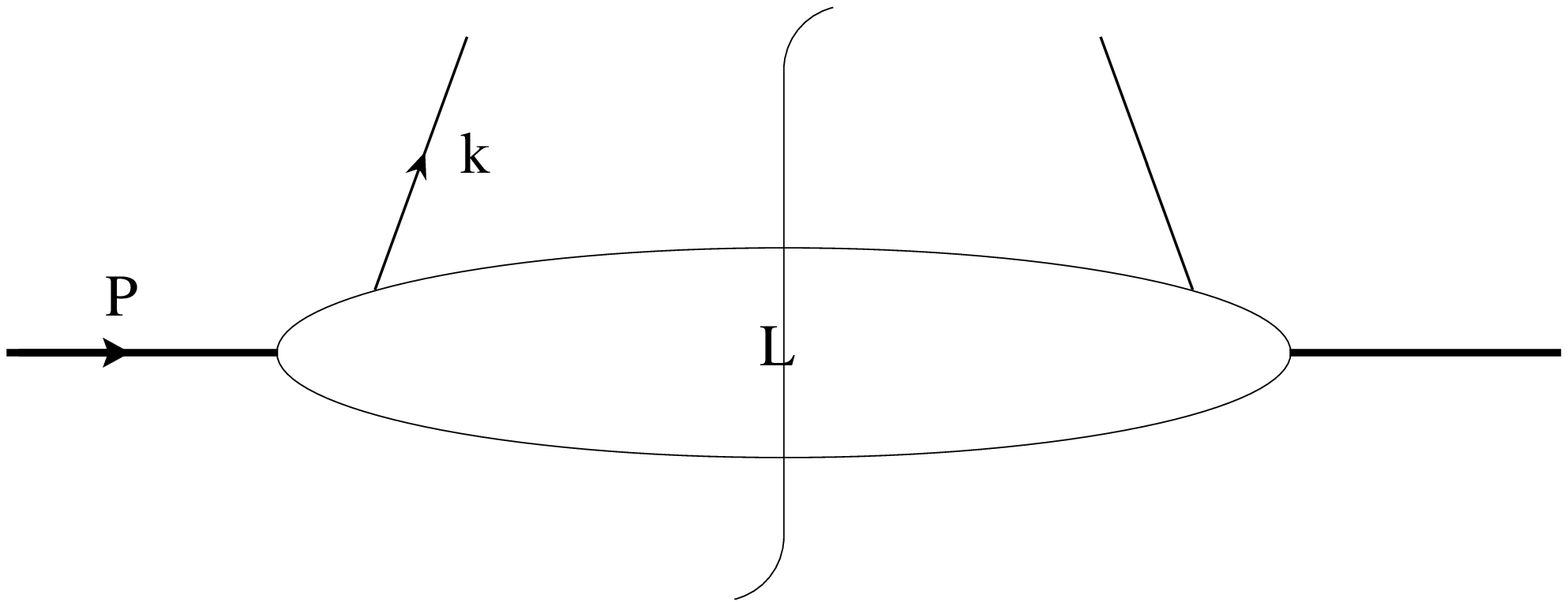}
   \caption{Two-parton correlation function used in definition of
         parton distribution function. } 
   \label{fig:pdf}
}

Since the cross section being computed is exclusive rather than
inclusive, the reasoning leading to the factorization theorem for
inclusive processes does not directly apply, and more general
arguments are mandatory \cite{JCC,phi3}.  Now a pdf
$f_i(x)$ is essentially the number density of partons of flavor $i$
and fractional longitudinal momentum $x$ with an integral over
transverse momentum and virtuality.  An examination \cite{JCC} of the
derivation of the event generator algorithms shows that the form of
this integral and its cutoff at large transverse momentum and
virtuality are determined by the showering algorithm.  For example,
if the Bengtsson-Sj{\"o}strand algorithm of Ref.\ \cite{BS} is used for
the kinematics of the initial-state parton, then the pdf is 
\begin{equation}
  \label{eq:pdf.oper.def}
  f_i(x) = \int d^4k \, \delta(x-k\cdot(k+q)/p\cdot q)
               \, \theta(Q^2-|k^2|)
               \, L(k,p),
\end{equation}
where $L(k,p)$ is a two-particle correlation function of two partons
in the target hadron, Fig.\ \ref{fig:pdf}.  The delta function gives the
definition of the longitudinal fractional momentum variable in the
algorithm, and the theta function implements the upper cutoff on
virtuality.  Some details concerning gauge invariance have not been
precisely specified, but for our purposes this will not matter.  In
any case the definition is different from the \MSbar{} definition, and
the specification of the showering algorithm implies a specific
prescription for defining the pdfs, without any further
choice.

In an event generator, the definitions of the parton kinematics in an
initial-state shower are phenomenologically manifested in the
kinematics of the hadronic final state.  Hence if the parton
kinematics are mismatched between the parton-shower algorithm and the
definition of the pdf, the calculation of the final state
is incorrect.  This contrasts with the calculation of an inclusive
cross section, where the relevant part of the final state is summed
over; all that matters in an inclusive cross section is that given the 
kinematics for the struck parton a final state is generated with
probability unity.  

As usual, at lowest order it is legitimate to approximate the pdfs in the
correct scheme by those in some other conveniently chosen scheme
(e.g., \MSbar{}).  This is because the error caused by the incorrect
pdf is of the same order as the error caused by the unimplemented NLO
correction in the hard scattering.  But beyond LO, this is not an
appropriate approximation.

The methods of, for example, P{\"o}tter \cite{Poetter}, suggest that the
scheme for the pdfs can be chosen at will, with the scheme dependence
of the pdfs being compensated for in the computation of the hard
scattering.  This reasoning appears to be incorrect.  It starts from
the assertion that the cross section is given by (`bare') pdfs
convoluted with on-shell partonic matrix elements computed
without any subtractions.  Although this statement is often repeated
in the literature, we know of no proof.  
Indeed it is a clearly unphysical
statement, since in the real world of QCD partons are never on-shell.
Moreover it is not necessary \cite{fact} for a correct proof of
factorization.  The incorrect starting assumption is particularly
inappropriate for work with an event generator, where one explicitly
treats the showering of partons that have much lower virtuality than
that largest scale in the process.  Furthermore, as we stated above,
the standard factorization theorem and its derivation are not
sufficient by themselves to derive an algorithm for parton showering,
at NLO accuracy.

Moreover, the method of P{\"o}tter results in the real-emission part of
the NLO hard scattering being obtained from {\em unsubtracted}
partonic scattering graphs integrated down to a small cutoff $s_{\rm
  min}$.  Real emission at NLO below the cutoff is assigned to the
same parton configuration as the LO term; for this to be a useful
approximation, $s_{\rm min}$ must be substantially smaller than the
primary scale $Q^2$ of the hard scattering.  This immediately implies
that there is a double logarithm of the ratio $s_{\rm min}/Q^2$ in the
integrated hard scattering coefficient.  Since there will be
corresponding logarithms in higher orders, this implies that the NLO
hard scattering in this method is an inappropriate way of truncating a
perturbation expansion.

Similar remarks apply to other proposals along similar lines, for
example that of Dobbs \cite{Dobbs}.

\subsection{Pdfs in MC event generators}
In \cite{JCC}, a subtraction method was introduced to consistently
take into account the LO and the NLO contributions for DIS in PYTHIA.
Two different algorithms were discussed and formulas for the
appropriate pdfs in terms of \MSbar{} pdfs were derived at the level
of gluon-induced NLO terms.  The gluon-induced term is particularly
important because of the large size of the gluon distribution 
functions at small
$x$.  The algorithms are the standard Bengtsson-Sj{\"o}strand algorithm
\cite{BS} and a modified algorithm of Collins \cite{JCC} which has
improved kinematic properties.  We label these algorithms ``BS'' and
``JCC''.

In this section we review the derivation, which is made by comparing
calculations of the contribution of quark $a$ to $F_2$ in the \MSbar{}
scheme with the corresponding calculation in each of the
event-generator algorithms:
\begin{eqnarray}
\label{F2.MSbar}
  F_2^a(x,Q^2) &=& x f_a^{\rm (\MSbar)}(x, \mu^2) 
\nonumber \\
        && +    \frac{ \alpha_s(\mu^2) }{ 2\pi }
       \int_x^1 d\xi  
       \frac{x}{\xi} f^{(\MSbar)}_g(\xi, \mu^2) 
        \left[ P(z) \ln \frac{Q^2(1-z)}{\mu^2z} -\frac{1}{2} + 4z(1-z)\right]
\nonumber \\
        && +\mbox{ NLO quark terms} + O(\alpha_s^2)
\\
\label{F2.BS}
        &=&  x f^{\rm BS}_a(x, Q^2)
      + \frac{ \alpha_s(Q^2) }{ 2\pi }
       \int_x^1 d\xi  \int_{-1}^1 d\cos\theta \frac{x}{\xi} 
        f_g(\xi, Q^2) \times
\nonumber \\
       && ~ ~ \times \frac{1}{1-\cos\theta}  
       \left\{
          \left[ P(z) - C(-\hat{t}) \frac{f_a(x)}{f_a(x_1)}
           P\left(\frac{x_1}{\xi}\right)
          \right]
          -\frac{1}{4} + \frac{3}{2}z(1-z)
       \right\}
\nonumber \\
        && + \mbox{ NLO quark terms} + O(\alpha_s^2)
\\
\label{F2.JCC}
        &=&  x f_a^{\rm JCC}(x, Q^2) 
\nonumber \\
    && +    \frac{ \alpha_s(Q^2) }{ 2\pi }
       \int_x^1 d\xi  
       \frac{x}{\xi} f_g(\xi, Q^2) 
        \left[ P(z) \ln 1/z  -\frac{1}{2} + 3z(1-z)\right]
\nonumber \\
        && + \mbox{ NLO quark terms} + O(\alpha_s^2) .
\end{eqnarray}
Here $z = x/\xi$ while $P(z) = \frac{1}{2} [z^2+(1-z)^2]$ is the
splitting kernel for gluon $\to$ quark-antiquark pair.
In the formula for the BS algorithm,  $x_1=x-\frac{1}{2}\xi(1-\cos\theta)$ and 
$-\hat{t} = Q^2(1-\cos\theta) \xi/2x$, while $C(-\hat{t})=\theta(Q^2+\hat{t})$ is
the cutoff function for the showering.

We define $F_{2,{\rm \MSbar}}^{\rm LO}$, $F_{2,{\rm BS}}^{\rm LO}$ and
$F_{2,{\rm JCC}}^{\rm LO}$ to be the first terms on the right of each of Eqs.\ 
(\ref{F2.MSbar}), (\ref{F2.BS}) and (\ref{F2.JCC}) respectively.
Similarly the second terms are called $F_{2,{\rm \MSbar}}^{\rm NLO}$,
$F_{2,{\rm BS}}^{\rm NLO}$ and $F_{2,{\rm JCC}}^{\rm NLO}$, respectively.

Formulas follow immediately \cite{JCC,Yujun} for the quark distribution 
function in the BS scheme and the JCC scheme in terms of those in the
\MSbar{} scheme:
\begin{eqnarray}
\label{pdf.JCC}
  x f^{\rm JCC}_a(x, \mu^2)
  &=&  x f^{(\MSbar)}_a(x, \mu^2)
\nonumber\\
   && + \frac{ \alpha_s(\mu^2) }{ 2\pi }
       \int_x^1 d\xi  
       \frac{x}{\xi} f^{(\MSbar)}_g(\xi, \mu^2) 
        \left[
          P(z) \ln (1-z) + z (1-z)
       \right]
\nonumber\\
   && 
   + \mbox{ NLO quark terms} + O(\alpha_s^2)
\\\nonumber
  &=&  x f^{(\MSbar)}_a(x, \mu^2) + F_{2,{\rm \MSbar}}^{\rm NLO} -
  F_{2,{\rm JCC}}^{\rm NLO}
   + \mbox{ NLO quark terms} + O(\alpha_s^2).
\end{eqnarray}
\begin{eqnarray}
\label{pdf.BS}
  x f^{\rm BS}_a(x, \mu^2)
  &=&  x f^{(\MSbar)}_a(x, \mu^2)
\nonumber\\
   && + \frac{ \alpha_s(\mu^2) }{ 2\pi }
       \int_x^1 d\xi  
       \frac{x}{\xi} f^{(\MSbar)}_g(\xi, \mu^2) 
        \Bigg \{
          P(z) \ln \frac{1-z}{z} + z (1-z) 
\nonumber \\
   && ~ ~ ~ \left.
       - \int_{-1}^1 \frac{d \cos\theta}{1-\cos\theta}  
        \left [ P(z) - C(-\hat{t}) \frac{f_a^{\rm BS}(x)}{f_a^{\rm BS}(x_1)}
        P\left(\frac{x_1}{\xi}\right)\right]
      \right\}
\nonumber\\
   && 
   + \mbox{ NLO quark terms} + O(\alpha_s^2),
\\
  &=&  x f^{(\MSbar)}_a(x, \mu^2) +F_{2,{\rm \MSbar}}^{\rm NLO} -
    F_{2,{\rm BS}}^{\rm NLO}
   + \mbox{ NLO quark terms} + O(\alpha_s^2).
\nonumber
\end{eqnarray}
Note that  Eq.\ (\ref{pdf.BS}) is a nonlinear equation in terms of
$f_a^{\rm BS}$; this arises from the particular treatment of
parton kinematics in the BS algorithm. When we calculate the numerical
value of $f_a^{\rm BS}$, we will use 
$f_a^{\rm \MSbar}$ in the integrand. We will justify this simplification in 
Sec.\ \ref{sec:consist}.  

\FIGURE{

\centering
\includegraphics [scale=0.43]{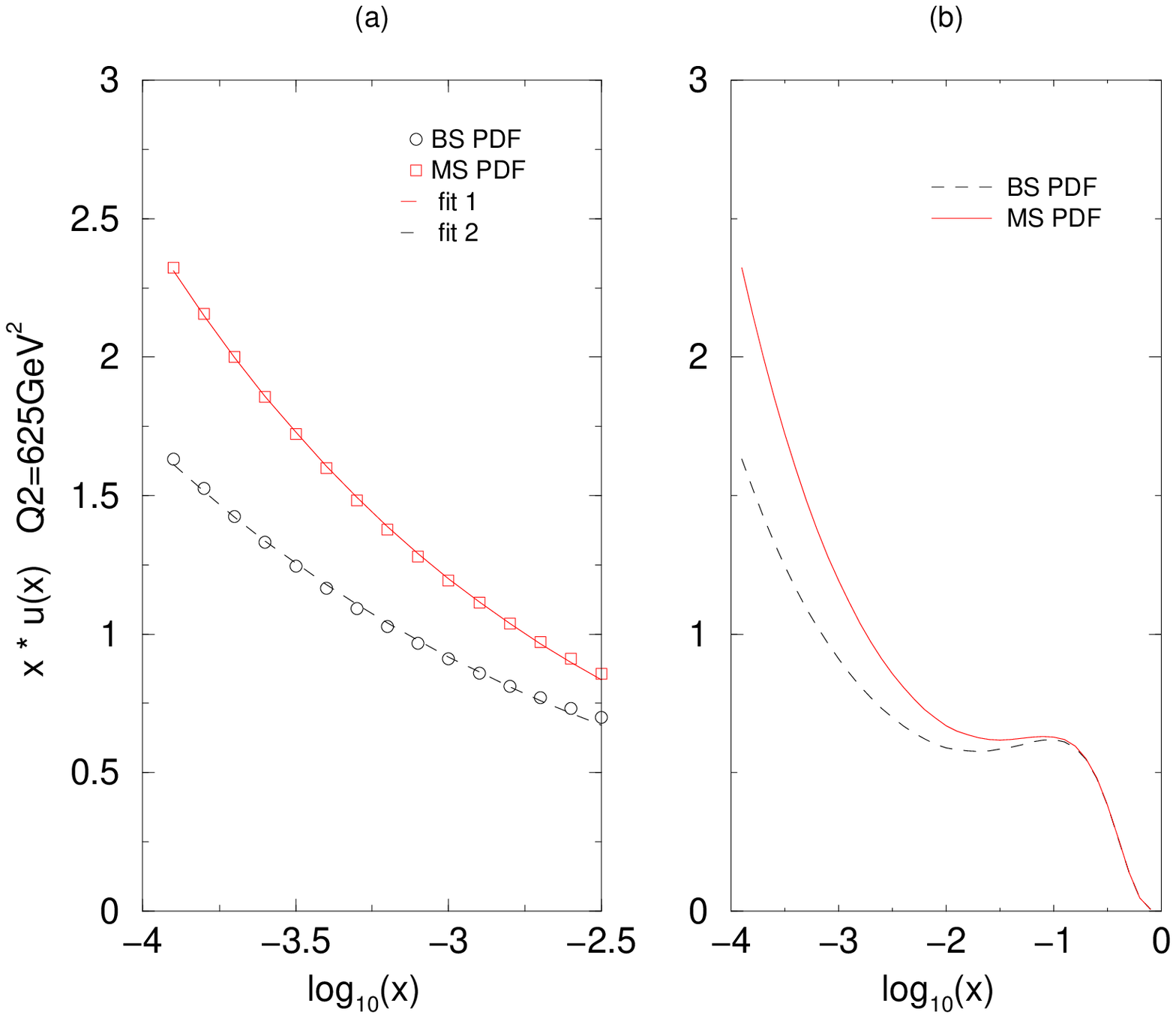}
\caption{The left-hand graph gives the $u$-quark distribution at small
  $x$ at $Q^2=625\,\GeV^2$. The lines are simple power law fits to the
  curves: $xu(x) = 0.135x^{-0.32} $ and $xu(x) = 0.14x^{-0.27}$ for
  the upper (\MSbar) and lower (BS scheme) curves. The inset gives the
  pdf at all $x$.}
\label{fig:pdffit}
}       

We have performed numerical calculations of the quark densities.  Our
code is based on earlier work by Sabine Schilling \cite{sabine}.  We
have made the code available at \cite{code}.  Some results are shown
in Fig.  \ref{fig:pdffit}, which gives the $u$ quark distribution
function in the $\MSbar$ and BS schemes at $Q^2=625\,\GeV^2$, with the
\MSbar{} density being that of CTEQ5 \cite{CTEQ5}.  This figure
clearly shows a large difference between the BS pdf and the \MSbar{}
pdf at small $x$.  The curves for the $d$-quark distribution function
are quite similar, so we do not show them.  In Secs.\ 
\ref{sec:results} and \ref{sec:consist}, we will analyze the scheme
dependence of the pdfs in more detail.

If we use $\MSbar$ pdfs rather than the ones appropriate to the BS
algorithm of the event generator, the exclusive cross section will be
in error by $\sim40\%$ at small $x$, although it is possible 
to get the correct
inclusive cross section, with the use of the well-known NLO correction
in this scheme.  As we will explain, this correction is unusually
small, so that good results can be obtained for the inclusive cross
section, i.e., for $F_2$, even without the use of the NLO correction,
if the \MSbar{} scheme is used.

We conclude that, while $\MSbar$ pdfs are well-defined quantities and
are appropriately used in calculations of inclusive cross sections,
they are not suitable for use in MC event generators where the fully
exclusive cross sections are our main concern.

        For the same reason, the pdfs used in \cite{Dobbs,Poetter} which are
essentially $\MSbar$ pdfs with ``scheme-dependence'' corrections, are not 
appropriate pdfs to use in the event generators.

\subsection{The NLO contributions to $F_2$ } 
\label{sec:results}
In this section we investigate the relative size of the NLO and LO
contributions to $F_2$, with the aid of some useful approximations,
and we show that the substantial differences between the schemes are
to be expected, because of the large size of the gluon distribution 
function.
A surprising result is that the NLO corrections to $F_2$ in the
\MSbar{} scheme are unusually small, as the result of special
cancellations.

From Eqs.\ (\ref{F2.MSbar}), (\ref{F2.BS}) and (\ref{F2.JCC}),
the relative sizes of the NLO terms can be estimated as follows:
\begin{eqnarray}
        \frac{F_2^{\rm NLO}}{F_2^{\rm LO}} &=& \frac{\alpha_s}{2\pi} 
        \int_x^1 d\xi    \frac{x}{\xi} f_g(\xi) O(1)
\nonumber \\
\label{f2.approx}
        &\sim& \frac{\alpha_s}{2\pi} f_g(2x) /f_a(x).
\end{eqnarray}
This estimate is appropriate to the small $x$ region.  We have first
observed that each integral contains a factor $\alpha_s/2\pi$, a factor of
the gluon distribution function and a factor of order unity.  At small $x$, the
important values of $\xi$ range from $x$ to a modest factor times $x$,
so that we set the argument of the gluon distribution function  to $2x$, as an
estimate of the typical value of $\xi$.

The ratio of NLO to LO would generally be at most of order
$\alpha_s/2\pi$, which is appropriate for a generic NLO correction, were it
not that the gluon distribution function is large at small $x$.
Clearly, there should be an enhancement of the NLO contribution, and
the above formula gives the expected size.  

Fig.\ \ref{fig:pdf.value} displays the numerical value of the ratio of
the NLO and the LO terms in $F_2$ for different factorization schemes
at $Q^2 = 33.8\,\GeV^2$, in the case of the $u$ quark.  The figure
also show the results for the simple estimate (\ref{f2.approx}).
Results for other flavors would be similar.

For the BS scheme, the large gluon distribution function at small $x$ does
indeed give a substantial effect: the NLO contribution is close to
$100\%$ of the LO contribution at $x=10^{-4}$.  The effect is much
smaller at large $x$.  The plot of the simple estimate
(\ref{f2.approx}) shows that this behavior is completely expected.
The accuracy of the approximation is an accident, but the overall size
is not.

\FIGURE{
\centering
\includegraphics[scale=0.4]{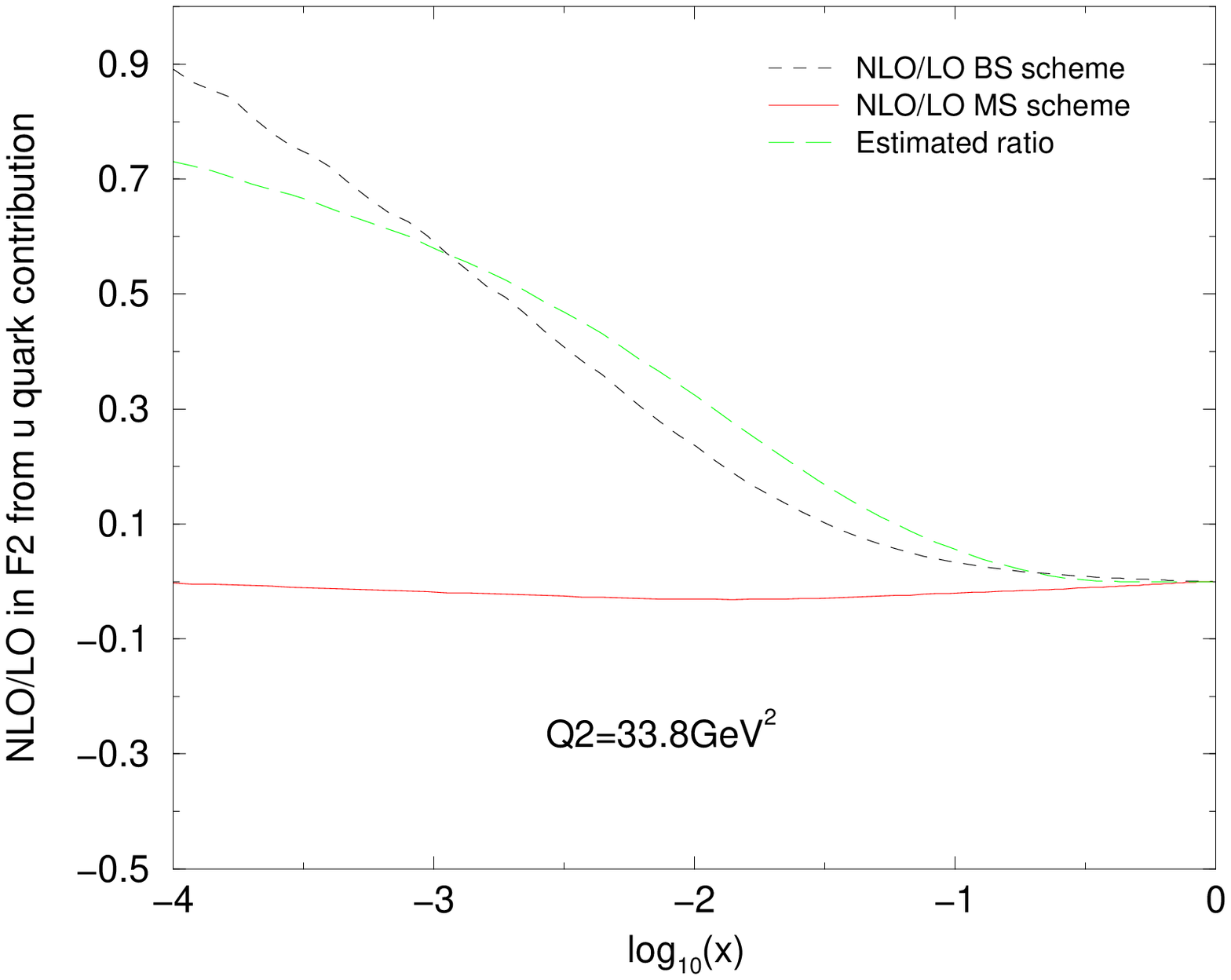}
\caption{Ratio of the NLO and LO terms in $F_2$ from u quark contribution
        for different factorization schemes.  We also show the simple estimate 
        $\frac{\alpha_s}{2\pi}  \frac{f_g(2x)}{f_a(x)}$ 
        Eq.\ (\ref{f2.approx}).}
\label{fig:pdf.value}
}

For the commonly used \MSbar{} scheme, the NLO corrections are rather
small for all $x$.  In view of the expected size of generic NLO
corrections, we should regard the small size of the correction in
\MSbar{} as an accident that is useful for phenomenology rather than
as a fundamental expectation.  The cancellation is associated with the
particular sizes of the positive and negative terms in the second line
of Eq.\ (\ref{F2.MSbar}).

To understand this cancellation better, it is convenient to perform a
slightly different approximation where we replace the pdfs
by power laws.  At small $x$, the gluon distribution function $f_g(x)$ 
is roughly $C/x^q$ and the quark distribution function $f_a(x) $ is 
roughly $C'/x^r$.  This gives the following approximations for each of the 
$F_2^{\rm NLO}$s:
\begin{eqnarray}
\label{nlo.MSbar}
  F_{2,{\rm \MSbar}}^{\rm NLO}(x, Q^2) & \approx & 
    \frac{ \alpha_s(Q^2) }{ 2\pi } C x^{1-q} \int_0^1 dz z^{q-1}
        \left[ P(z) \ln \frac{1}{z}  + 4z(1-z)\right]
\nonumber \\
   && - \frac{ \alpha_s(Q^2)}{2\pi} C x^{1-q} \int_0^1 dz z^{q-1}
        \left[ P(z) \ln \frac{1}{1-z}  + \frac{1}{2} \right]
,
\\
\label{nlo.BS}
    F_{2,{\rm BS}}^{\rm NLO}(x, Q^2) & \approx & 
    \frac{ \alpha_s(Q^2) }{ 2\pi } C x^{1-q} \int_0^1 dz z^{q-1}
       \Bigg\{ P(z) \ln \frac{1}{z} -\frac{1}{2}+3z(1-z) 
\nonumber \\
    &&+ \int_O^{2z} dy 
       \left[P(z) \frac{1-(1-\frac{y}{2z})^r}{y}
             - \left(1-\frac{y}{2z}\right)^r
               \left( \frac{1}{4}y-z+\frac{1}{2} \right)
       \right]
    \Bigg\}
\nonumber \\
    & \approx & \frac{ \alpha_s(Q^2) }{ 2\pi } 
        C x^{1-q} \int_0^1 dz z^{q-1}
       \left[ P(z) \ln \frac{1}{z} +\frac{3}{2}z(1-z)+\frac{1}{3}z^2\right]
, 
\\
\label{nlo.JCC}
    F_{2,{\rm JCC}}^{\rm NLO}(x, Q^2) & \approx & 
    \frac{ \alpha_s(Q^2) }{ 2\pi } C x^{1-q} \int_0^1 dz z^{q-1}
        \left[ P(z) \ln \frac{1}{z} +3z(1-z)-\frac{1}{2}\right]
.
\end{eqnarray}  
In each case, the lower limit of the $z$ integral can be set to zero
when $x$ is 
small, and we therefore have a simple power of $x$ times a fixed
integral over $z$.

The integrals of $z$ in the above equations are basically all of
$O(1)$.  The exponent $q$ is about $1.2\sim 1.4$ for 
the gluon distribution function.
The exponent $r$ for the quark distribution is close to 1 and less than
$q$.  Therefore, the NLO corrections will be enhanced at small $x$.
However there are negative terms in the \MSbar{} integral which
results in a cancellation.  There is a weaker cancellation in the
integral for the JCC scheme, but there is no cancellation in the
integral for the BS scheme.

\subsection{Accuracy of nonlinear term in BS pdfs} 
\label{sec:consist}

The BS quark distribution function is related to the $\MSbar$ distribution 
function by the
nonlinear integral shown in Eq.\ (\ref{pdf.BS}).  As is normal, we
replace the BS pdfs on the right-hand side by the \MSbar{} pdfs, so
that we get a formula involving \MSbar{} pdfs only.
Generally this is the normal procedure since the change involves an
effect of relative order $\alpha_s$ in an NLO term.  In transformations
that are linear in the pdfs, this is quite sensible: All the
errors are handled by the uncalculated terms of yet higher order.
However, this is more delicate for the nonlinear formula,
particularly given that the quark distribution functions have 
large corrections.  A
linear formula, as for the JCC scheme, only needs the 
gluon distribution function,
but our nonlinear formula also involves the quark distribution functions.

In this section we will show that this issue does not in fact affect
the accuracy of our calculations, since the right-hand side of Eq.\ 
(\ref{pdf.BS}) only involves a {\em ratio} of quark distribution functions; the
ratio of BS quark distribution functions, 
$f_a^{\rm BS}(x)/f_a^{\rm BS}(x_1)$, is
equal to the ratio of the \MSbar{} distribution functions to a 
good approximation.

Our demonstration is semi-analytic, so that we can see that the result 
is robust against changes in the pdfs. 

In Fig.\ \ref{fig:pdffit}, we fit the small $x$ pdfs with $y=Ax^B$.
The exponent $B$ of $x$ depends on $Q^2$, but the difference of
exponent between the BS scheme and the $\MSbar$ scheme is roughly the
same for all $Q^2$ and approximately equal to 0.05.

The error introduced by the replacement of the BS quark distribution
functions by the 
\MSbar{} distribution functions on the right-hand side of Eq.\ (\ref{pdf.BS})
 is then
\begin{eqnarray}
\label{delta}
   \delta & \equiv & 
    \frac{ \alpha_s(Q^2) }{ 2\pi } \int_0^1 dz \frac{x}{z} 
        f_g\left(\frac{x}{z}\right)
        \int_{1-2z}^1 \frac{d \cos \theta}{1-\cos\theta}
        P(z_1) \left[ 
        \frac{f_a^{\rm \MSbar}(x)}{f_a^{\rm \MSbar}(x_1)} -
        \frac{f_a^{\rm BS}(x)}{f_a^{\rm BS}(x_1)} \right ] ,
\end{eqnarray}  
where $z_1= z- (1-\cos\theta)/2$.

The error $\delta$ is the largest in small $x$ region because of the large
difference between the $\MSbar$ pdfs and the BS pdfs.  When $\cos\theta \to 1-2z$,
$x_1 \to 0$, and then $ f_a(x) / f_a(x_1) \to 0$, therefore $\delta$ is very
small.  When $\cos\theta \to 1$, $x_1 \sim x$, we have
\begin{equation}
        \frac{f_a(x)}{f_a(x_1)} \approx 1 - \frac{\xi}{2}(1-\cos\theta) 
        \frac{f'_a(x)}{f_a(x)}
\end{equation}
and $f_a^{\MSbar}(x) \sim A_1/x^{1.32}$, $ f_a^{\rm BS}(x) \sim A_2/x^{1.27}$.

        We can see that $\delta$ depends on the pdfs through their difference
in exponents of $x$, rather than on the actual value of pdfs. Given
the exponents of the BS pdf and the $\MSbar$ pdf, we have,
\begin{equation}
        \delta \approx \frac{\alpha_s}{2\pi} \int_0^1 dz
        \frac{x}{z} f_g(\frac{x}{z}) P(z) / z \times 0.025
\end{equation}
which is less than $5\%$ of $F_{2,{\rm BS}}^{\rm NLO}$.
Therefore, our simplification is valid up to 
NLO accuracy.

\section{Conclusion}

We explained that the pdfs in MC event generators are determined by
the showering algorithm, and cannot be freely chosen, unlike the case
for pdfs used in inclusive calculations. The rules for calculating
the hard-scattering coefficients at higher orders are then
unambiguously defined.  We then presented some numerical calculations
of the quark distribution functions to be used with the BS algorithm.  At 
small $x$
the corrections are large, so that the commonly used $\MSbar$ pdfs are
inappropriate for use in event generators.  We used some simple
approximations to understand the size of the corrections and to show
that the large correction is to be expected. 

The code for the MC-specific pdfs is available at \cite{code}.

\section*{Note Added}

Early papers on HERWIG, for example the paper of Marchesini and Webber
\cite{MW}, also mentioned the idea that the showering algorithm
entails a particular definition of the pdfs, with cutoffs on parton
kinematics that correspond to cutoffs in the showering.  However,
since the event generator was only implemented at leading order, the
need for modified pdfs was not emphasized in Ref.\ \cite{MW}; it was
sufficient to use unmodified pdfs from standard fits.

We thank the referee for bringing this work to our attention.

\section*{Acknowledgements}
        We would like to thank Hannes Jung and Sabine Schilling 
for discussions and assistance. We would also like to thank DESY and the 
II Institut f{\"u}r Theoretische Physik der Universit{\"a}t Hamburg 
for their hospitality during the starting of this work. This work
was supported in part by the U.S.Department of Energy under grant number 
DE-FG02-90ER-40577.


\end{document}